\documentclass[aip,reprint]{revtex4-1}
\usepackage[T1]{fontenc}
\usepackage[latin9]{inputenc}
\usepackage{textcomp}
\usepackage{amstext}
\usepackage{graphicx}

\begin{document}

\preprint{This line only printed with preprint option}

\title{Giant coercivity of dense nanostructured spark plasma sintered barium
hexaferrite}

\author{F. Mazaleyrat}
\email{mazaleyrat@satie.ens-cachan.fr}
\homepage{http://www.satie.ens-cachan.fr}
\affiliation{SATIE, ENS Cachan, CNRS, Universud, 61 av President Wilson, F-94235,
Cachan, France}

\author{A. Pasko}
\affiliation{SATIE, ENS Cachan, CNRS, Universud, 61 av President Wilson, F-94235,
Cachan, France}

\author{A. Bartok}
\thanks{Presently at LGEP, Universite Paris Sud, Orsay.}
\affiliation{SATIE, ENS Cachan, CNRS, Universud, 61 av President Wilson, F-94235,
Cachan, France}

\author{M. LoBue}
\affiliation{SATIE, ENS Cachan, CNRS, Universud, 61 av President Wilson, F-94235,
Cachan, France}

\date{\today}

\begin{abstract}
Enhencement of
the coercivity of ferrite magnets in order to make them true hard
magnets -- i.e. to get a coercive field higher than the residual magnetization-- is still a very important issue due to the limited ressource in rare-earth.  Thus, an alternative can be found in making very fine grain ferrite
magnets but it is usually impossible to get small grains and dense
material together. In this paper, it is shown that the spark plasma
sintering method (SPS) is able to produce close to 80\% dense material
with crystallites smaller than 100 nm. The as prepared bulk sintered
anisotropic magnets exibits coercive field of 0.5 T which is close
to 60\% of the theoretical limit and only few \% below that of loose
nano-powders. As a result, the magnets behave nearly ideally (-1.18
slope in the BH plane second quandrant) and the energy product reaches
8.8 kJm$^{-3}$, the highest value achieved in isotropic ferrite
magnet to our knowledge.
\end{abstract}

\maketitle

\section{introduction}

In the last years, much efforts have been devoted to the enhancement
of the coercivity of hexagonal M-ferrites. The most fructuous results
have been obtained by substitution of La and Co in strontium ferrite
according to the formula Sr$_{1-x}$La$_{x}$Fe$_{12-y}$Co$_{y}$O$_{19}$
\cite{Tenaud2004}. A very good compromize have been found for equi-molar
La and Co compound with $x=y=0.2$. Under optimal processing and
sintering conditions, anisotropic magnets have been obtained with
a very good rectangularity and an intrinsic coercivity as high as
334 kAm$^{-1}$(viz. $\mu_{0}H_{CJ}=0.42$~T) which represents a
20\% improvement compared to regular anisotropic Sr magnets. This very
good result, however, has to be compared with  the anisotropy
field $\mu_{0}H_{K}=2.27$~T of this compound, showing that, alltogether,
the coercivity is only a fifth of the anisotropy field. This effect,
known as Brown's paradox, is related to magnetic domain structure.
On the other hand, a strong enhencement of the coercivity can be obtained
by reducing the grain size below the single domain limit as it was
proposed by N\'eel in 1942 (published ony after World War 2 \cite{Neel1,Neel2}).
According to this theory --usually referred under the name of Stoner
\& Wohlfarth \cite{SW}-- the upper limit of the coercivity is $H_{K}$
for perfectly oriented and $0.48\times H_{K}$ for an assembly of
non-interacting uniaxial particles with isotropic distribution of
easy axes. Several authors have succeed to produce BaFe$_{12}$O$_{19}$
nanoparticles by soft chemical route and they have found coercivities
as high as 0.58~T for a mean grain size about 100~ nm \cite{Pankhurst1996,zhong1997}.
As the anisotropy field is 1.7~T, the theoretical value for an isotropic
magnet is 0.82~T, so that they reached 70\% of the upper limit, showing
how promising should be this route. There is however a big drawback
with this approach: these results have been obtained with loose powders
only, whereas in practice, it is necessary to supply solid magnets.
On the one hand, embeding in resin can keep this high coercivity but reduce
dramatically the residual induction (due to dilution effect) in such
a way that the energy product is not improved. On the other hand,
sintering produce grain growth and the coercivity drops down to about
0.2~T.  In this paper, we are showing that, starting from hexaferrite nanopowder, Spark  Plasma Sintering (SPS)
allows to produce nonostructured Ba-ferrite with a density close to 90\% with
almost no reduction in coercivity by opposition to SPS synthesis of nanosize Ba-ferrite in a single step \cite{zhao2006,zhao2008evidence}. 
\begin{table*}
\begin{centering}
\begin{tabular}{cccccccccc}
\hline 
{ Sample} & { Sintering} & \multicolumn{3}{c}{{ Lattice constants}} & { Cr. size} & { Density} & \multicolumn{3}{c}{Hysteresis parameters}\tabularnewline
\cline{3-5} \cline{8-10} 
{\small Ref.} & {\small time (min)} & {\small $a$ (nm)} & {\small $c$ (nm)} & {\small $v$ (nm$^{3}$)} & {\small $D$ (nm)} & {\small{} (\%)} & $m_{S}$(Am$^{2}$kg$^{-1}$) & $m{}_{R}/m_{S}$ & $\mu_{0}H_{C}$(T)\tabularnewline
\hline
Lit* & single cryst. & 0.58920 & 2.31830 & 0.69699 & -- & 100 &  &  & \tabularnewline
Com\dag & coarse grain & 0.58928 & 2.32010 & 0.69772 & >1000 & >95 & 62 & 0.51 & 0.26\tabularnewline
CP & as calcinated & 0.58946 & 2.32185 & 0.69867 & 60 & -- & 55 & 0.51 & 0.56\tabularnewline
SPS1 & 0 & 0.58878 & 2.32378 & 0.69764 & 70 & 76 & 66 & 0.57 & 0.51\tabularnewline
SPS3 & 13 & 0.58858 & 2.32224 & 0.69670 & 84 & 88 & 66 & 0.57 & 0.49\tabularnewline
SPS4 & 20 & 0.58892 & 2.32259 & 0.69761 & 77 & 86 & 62 & 0.58 & 0.49\tabularnewline
\hline
\end{tabular} 
\par\end{centering}

\caption{\label{tab:Structural-parameters}Structural and magnetic parameters of calcinated
powder and SPS samples. Accuracy is $\pm$5 nm for  grain size (from XRD) and $\pm$1\% for density. *Litterature values are given for reference; \dag refers to commercial magnet.}

\end{table*}

\section{Experimental details}

\begin{figure}
 \centering
 \begin{center}
 \includegraphics[width=0.7\columnwidth]{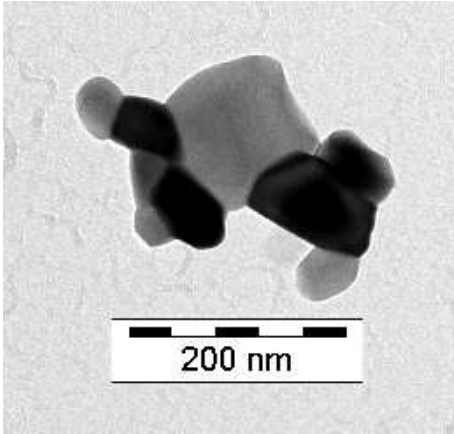}
\end{center}

 \caption{Typical hexaferrite particle as received after clacination}
 \label{fig: TEM}
\end{figure}

Barium hexaferrite powders have been prepared by a sol-gel citrate
precursor method \cite{Licci84SolGel,srivastava87SG,yu06solgel}.
High purity iron(III) nitrate, barium hydroxide and citric acid were
used as starting materials with the molar ratio of citrate to metal
ions 2:1. Iron(III) nitrate was dissolved in deionized water and quantitatively
precipitated with excess of ammonia solution as iron(III) hydroxide.
The precipitate was filtered and washed with water until neutrality.
Then the obtained iron(III) hydroxide was dissolved in a vigorously
stirred citric acid solution at 60-70\textdegree{}C. Barium hydroxide
was added according to the desired composition. Polycondensation reaction
with ethylene glycol to prevent segregation \cite{Licci84SolGel}
was not used. Instead, pH value of the solution was adjusted to 6
for better chelation of metal ions \cite{yu06solgel}. Water was slowly
evaporated at 80-90\textdegree{}C with continuing stirring until a
highly viscous residue is formed. The gel was dried at 150-170\textdegree{}C
and heat treated at 450\textdegree{}C for 2 hours for total elimination
of organic matter. Finally, the inorganic precursor with homogeneous
cationic distribution was calcined at 900\textdegree{}C for 2 hours
with heating/cooling rate 200 K/hour to synthesize the barium hexaferrite
phase. The calcination temperature have been chosen high enough to
allow complete formation of the M phase and low enough to prevent
grain growth. A representative TEM image of grain size and morphology is given in Fig.\ref{fig: TEM}. Details of this optimization procedure will be given in a forthcoming paper.

The samples have been sintered in a Sumitomo Dr Sinter spark plasma sintering (SPS) machine under a pressure
of 50 MPa and neutral atmosphere in a graphite die. The heating rate
was 160~Kmin$^{-1}$, up to 800\textdegree{}C. This temperature
was kept for a duration between 0 and 20 min and then cooled at a
rate of 200 ~K.min$^{-1}$. 

XRD diffractograms have been recorded using a PANalytical X'Pert Pro 
equiped with a linear detector and the analysis of spectra was conducted
with the Rietveld based software MAUD \cite{MAUD}. The morphology
of sinterd sample was obsvered with a Hitachi SEM. Density was mesured
using standart Archimedes method and the hysteresis loops where recorded
at room temperature using a LakeShore vibrating sample magnetometer
with a maximum field of 2 T.

\section{Results and discussion}

\begin{figure}
\begin{centering}
\includegraphics[width=.8\columnwidth]{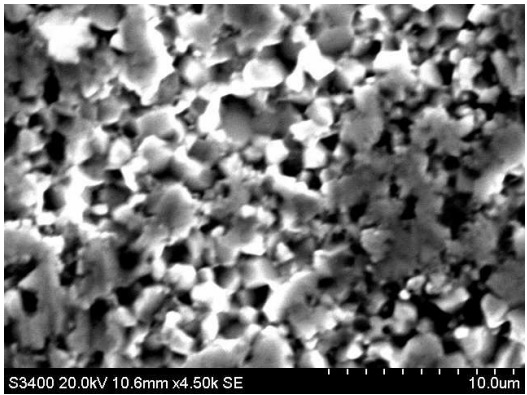}
\par\end{centering}

\caption{\label{fig:SEM}SEM picture of sample SPS4}

\end{figure}

X-ray diffractogram of the powder after calcination shows all characteristic
peaks of M-type hexaferrite (magnetoplumbite) with relatively broad
shape. The lattice parameters are very close to those of literature reference
samples and the average grain size was found to be about 60 nm (see
Table \ref{tab:Structural-parameters}). This value obtained by fitting diffractograms with MAUD is confirmed by TEM (see Fig.\ref{fig: TEM}). Although some particles are bigger than 100 nm, most of then remains close to 50 nm. A slight decrease in the
$a$ axis paramer has been observed together with a increase of the
$c$ axis parameter upon sintering but it doesn't seem to be significant
as the unit cell volume is almost constant. If this variation could be
considered as significant, one would attribute it to internal stresses
produced by the sintering process under uniaxial pressure. As expected
with the SPS process, the average crystallite size is not much enhanced
even after 20 min compared to the calcinated powder. This effect is
related to the high speed of the process which allows atomic short
range diffusion (involved in sintering process) but not long range
diffusion (involved in grain growth process). In contrast, the density
changes appreciably with time as it has been observed with most of
materials sintered by SPS. A non negligible porosity remains after
sintering as it is also seen in the SEM picture in Fig. \ref{fig:SEM}.
Only the sample SPS4 is shown but others exhibits very similar features.
From these pictures, it is concluded  that the material is composed
of approximatively 1 \textmu{}m grains composed of 70-80 nm crystallites
with $\sim$500 nm node pores.

\begin{figure}
\includegraphics[width=1\columnwidth]{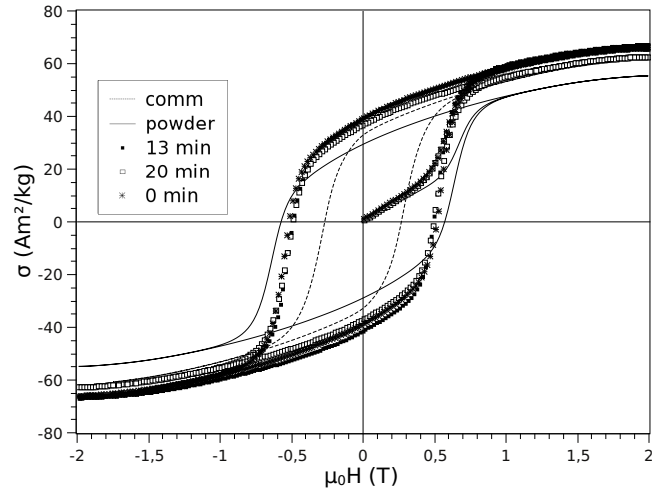}

\caption{\label{fig:VSM}Hysteresis (specic magnetization as a function of
applied induction field) loops of commercial, as-calcinated and SPS
sintered samples. }

\end{figure}

The hysteresis loops of the different samples are shown in Fig. \ref{fig:VSM}.
Comparison with conventional coarse-grain sample (commercial) immediatly
shows the interest of reducing the grain size down to nm range: the
sample measured after calcination in the form of loose powder exhibits
a coercive field of 0.56 T, which is twice that of the coarse grain
counterpart. In comparison with the theoretical field for an anisotropic
magnet, this value is 70\%, very close to results obtained by others
\cite{zhong1997,Pankhurst1996}. The rectangularity of the loop is
very close to 50\% as expected from N\'eel calculations for a uniform
distribution of easy axes. After SPS sintering, the density is already
high, even for a very short heating-cooling cycle. A marked increase
in saturation magnetization is observed, probably due to uncomplete
transformation of the precursors during calcination. However, samples
sintered for 0 to 13 min, still have sensitively higher value than
the coarse grain sample, and drop to the same value for longer treatment
time. This effect is probably due to an excess in oxygen gradually
released due to the slightly reducing conditions.

The most stricking effect of SPS processing is the high value of the coercivity.
Indeed,  sintering reduces the coercive field, by only several
tens of mT, so that coercivity values remain very close to 0.5\,T for all SPS samples.
This feature is due to the fact the crystallite size remains unchanged
clearly under 100\,nm, much below the single domain limit, $D_{SD}={36\mu_{0}\sqrt{AK_{1}}}/{J_{S}^{2}}=235\,\mathrm{nm}$
given after Kittel formula, where $K_{1}=338\,\mathrm{kJm^{-3}}$, $A=5\,\mathrm{pJm^{-1}}$ and
$J_{S}=0.5\,\mathrm{T}$. 
As a consequence, nucleation of domain walls is impossible and the magnetization reversal should be processed by rotation. Although it is still far from the theoretical limit for isotropic magnets $0.48H_{K}\approx0.85$\,T the coercivity is doubled compared to regular isotropic ferrite magnets. This feature is very important in applications, since it considerably improves the resistance upon demagnetization. From pratical point of view the characteristic in the BH plane is more relevant. Computation of the BH loop $B=\mu_{0}(\sigma\rho\delta+H)$, with $\sigma$ the specific magnetization, $\rho$ the bulk specific mass and $\delta$ the relative density, yields for the extrinsic coercivity $\mu_{0}H_{CB}=0.2$\,T, the remnant induction $B_{R}=0.23$\,T and the energy product $(BH)_{max}=8.9$\,kJm$^{3}$ compared with 0.15\,T, 0.22\,T and 7\,kJm$^{3}$ respectively for regualar sample.

\section{Conclusion}
It has been demonstrated that SPS technique is an efficient and powerful tool for the sintering of nanostructured isotropic ferrite magnets as the coercivity of the powder can be obtained in dense samples. The energy product have been improved by 30\% and the hardness against demagnetization by a factor 2. In addition  low temperature sintering meets the requirement of LTCC technology with no need of glass addition \cite{Liu_JAP2010}.

\bibliographystyle{aipnum4-1}
%

\begin{acknowledgments}
This work was partly supported by the EC FP7 project SSEEC under grant
number NMP-SL-2008-214864. A.B. greatly aknowledge ENS Cachan for a 6 month scolarship
in the frame of international scolarship program 2008/2009.

P. Audebert, Pr at ENS Cachan Chemical department PPSM-CNRS is greatly
acknowledged for his advices in sol-gel production. 
\end{acknowledgments}

\end{document}